\documentclass{PoS}
\usepackage{amsmath}
\usepackage{amssymb}
\usepackage{amsbsy}  
\usepackage{amsfonts}
\usepackage{graphicx}
\usepackage{subfigure}
\usepackage{latexsym}
\usepackage{mathrsfs}
\usepackage{dcolumn} 
\usepackage{wasysym}
\usepackage{url}
\newcommand{\beq}{\begin{equation}}
\newcommand{\eeq}{\end{equation}}

\newcommand{\be}{\begin{eqnarray}}
\newcommand{\ee}{\end{eqnarray}}
\title
{Quark mass, scale and volume dependence of
topological charge density correlator in Lattice QCD}
\ShortTitle{Quark mass, scale and volume dependence of
TCDC in Lattice QCD}
\author{Abhishek Chowdhury$^{a}$,
Asit K. De$^{a}$,
A. Harindranath$^{a}$,
Jyotirmoy Maiti$^{b}$, 
\speaker{Santanu Mondal}$^a$\\
\llap{$^{a}$}Theory Division, Saha Institute of Nuclear Physics \\
 1/AF Bidhan Nagar, Kolkata 700064, India\\
\llap{$^{b}$}Department of Physics, Barasat Government College,\\
10 KNC Road, Barasat, Kolkata 700124, India\\

E-mail: \email{santanu.mondal@saha.ac.in}}

\date{August 9, 2012}
\abstract{
We study the two-point Topological Charge Density 
Correlator (TCDC) in lattice QCD with two degenerate flavours of naive Wilson 
fermions and unimproved Wilson gauge action at two values of lattice spacings 
and different volumes, for a range of quark masses. Configurations are 
generated with DDHMC algorithm and smoothed with HYP smearing.
 In order to shed 
light on the mechanisms leading to the observed
suppression of topological susceptibility with respect to the decreasing quark mass
and decreasing volume,
in this work, we carry 
out a detailed study of the  two-point TCDC.
We have shown that, 
(1) the TCDC  is negative beyond a positive core and radius of the 
core shrinks 
as lattice spacing decreases, 
(2) as the volume decreases, the magnitude of the 
contact term and the 
radius of the positive core
decrease and the magnitude of the negative peak increases 
resulting in the 
suppression of the topological susceptibility as the volume decreases, 
(3) the contact term and radius of the positive core decrease with
decreasing quark mass at a given lattice spacing and  the
negative peak increases with decreasing quark mass resulting
in the suppression of the topological susceptibility with decreasing quark 
mass, 
(4)  increasing levels of smearing suppresses the contact term and
the negative 
peak keeping the susceptibility intact and 
(5) both the contact term and the negative peak diverge in nonintegrable 
fashion as lattice spacing decreases. 

}
\FullConference{The 30th International Symposium on Lattice Field Theory\\
                 June 24 - 29,  2012\\
                 Cairns, Australia}

\begin{document}
\maketitle
\section{Introduction}
As a consequence of the reflection positivity and the pseudoscalar nature of 
the relevant local operator in Euclidean quantum field theory, the two-point 
Topological
Charge Density Correlator (TCDC) is negative at arbitrary non-zero 
distances \cite{es}. In the continuum theory, close to the origin 
the two-point TCDC is negative and singular. From power
counting, the singularity $\sim -|x|^{-8}$ up to possible logarithms
and hence is non-integrable. In order to obtain a positive and finite 
space-time integral (susceptibility), the TCDC should have a positive 
non-integrable singularity at the origin \cite{es,as}.  
However, it is possible
to give a rigorous definition of topological susceptibility 
in Lattice QCD without 
power divergences using Ginsparg-Wilson fermion \cite{giusti,luscher1}.

As the authors of Ref. \cite{es} pointed out long 
time ago,  this has non-trivial consequences for the derivation and 
interpretation of the Witten-Veneziano expression \cite{wv} for the 
$\eta^{\prime}$ mass.  The negativity of the TCDC 
also has non-trivial consequences related to the nature of 
topological charge structure in QCD vacuum \cite{ih}. 

The issues related to two-point TCDC are best studied in the theory 
rigorously formulated on a Euclidean lattice. However,  the 
lattice theory defined by a particular action may not be reflection positive. 
Fortunately, this is not a concern for the Wilson fermion.
However, the breaking of chiral symmetry by Wilson term
may lead to uncancelled divergences in topological susceptibility.
Thus it is important to calculate topological susceptibility
with Wilson fermion to check whether the cancellation indeed happens
so that Wilson lattice QCD belongs to the same universality class
as continuum QCD.
The lattice operator for the topological charge density 
$q(x)$ may extend over several lattice spacings, 
and thus for sufficiently small $x$, 
the continuum like behaviors are not expected. Nevertheless,  
continuum  properties are expected to emerge as lattice spacings become 
smaller and smaller. Specifically, 
on a lattice with 
lattice  spacing $a$, TCDC remains positive within a radius $r_c$, 
which is expected to shrink to zero as $a \rightarrow 0$.  
The first investigation of lattice spacing dependence of 
the radius of the positive core and
the negativity beyond the positive core of TCDC in lattice 
QCD was carried out in Ref. \cite{ih2} in the context of overlap based 
topological charge density in quenched QCD. Later, similar study was carried 
out \cite{fb} for a variety of lattice QCD actions with and without 
quarks where discretization errors 
appear only at ${\cal O}(a^2)$.

Flavour singlet axial Ward-Takahashi identity relates the topological 
susceptibility $\chi$, which is the four-volume integral of TCDC, to the 
chiral condensate in the chiral limit \cite{rjc,ls}. As a consequence, $\chi$
vanishes linearly in the quark mass in the chiral limit. Furthermore, at 
a given value of the quark mass, $\chi$ is suppressed as volume decreases 
\cite{ls, sd}.  
As part of an on-going program \cite{p1,p2} to study the chiral properties of 
Wilson lattice QCD (unimproved fermion and gauge actions), recently, 
we have demonstrated the suppression of 
topological susceptibility with decreasing quark mass in the case of 
unimproved Wilson fermion and gauge action \cite{ac1,ac2} where, 
the suppression of $\chi$ with decreasing volume was also shown. 
In order to shed 
light on the mechanisms leading to these suppressions, in this work, we carry 
out a detailed study of the  two-point TCDC. 
\section{Measurements}


\begin{table}
\begin{center}
\begin{tabular}{|l|l|l|l|l|l|l|l|l|l|}

 \multicolumn{7}{c}{$\beta = 5.6$} \\
\hline
$tag$&$lattice$& $\kappa$& $block $&{$N_2$}& {$N_{trj}$}  &{$\tau$}\\ 
\hline
{$A_{2b}$}&{$16^3\times32$}&{$0.158$}&{$8^4$}&{$10$}&{$6816$}&{$0.5$} \\
\hline 
{$B_{1b}$}&{$~~~~~,,$}&{$0.1575$}&{$12^2\times6^2$}&{$18$}&{$13128$}
&{$0.5$} \\
{$B_{3b}$}&{$~~~~~,,$}&{$0.158$}&{$12^2\times6^2$}&{$18$}&{$13646$}&{$0.5$} \\
{$B_{4b}$}&{$~~~~~,,$}&{$0.158125$}&{$12^2\times6^2$}&{$18$}&{$11328$}
&{$0.5$}\\
{$B_{5b}$}&{$~~~~~,,$}&{$0.15825$}&{$12^2\times6^2$}&{$18$}&{$12820$}&{$0.5$}\\
\hline
{$C_2$}&{$32^3\times64$}&{$0.158$}&{$8^3\times16$}&{$8$}&{$7576$}&{$0.5$}\\
\hline \hline

  \multicolumn{7}{c}{$\beta = 5.8$} \\
\hline
$tag$&$lattice$& $\kappa$& $block$ &{$N_2$}& {$N_{trj}$}  &{$\tau$} \\
\hline
{$D_1$}&{$32^3\times64$}&{$0.1543$}&{$8^3\times16$}&{$8$}&{$9600$}&{$0.5$}\\
{$D_3$}&{$~~~~~,,$}&{$0.15462$}&{$8^3\times16$}&{$24$}&{$7776$}&{$0.5$} \\
\hline \hline
\end{tabular}
\end{center}
\vspace{-.5cm}
\caption{Lattice parameters and simulation statistics.
Here $block$, $N_2$, $N_{trj}$ and  $\tau$
refers to HMC block, step number for the force $F_2$, number of HMC 
trajectories and the Molecular Dynamics trajectory length respectively.   }
\label{table1}
\end{table}

We have generated ensembles of gauge configurations by means of DDHMC 
algorithm \cite{ddhmc} using unimproved Wilson fermion and gauge actions with 
$n_f=2$ mass degenerate quark flavours. At $\beta=5.6$ the lattice volumes are 
$16^3 \times 32$, $24^3 \times 48$ and $32^3 \times 64$ and the renormalized 
quark mass ranges between $25$ to $125$ MeV ($\overline{\rm MS}$ scheme 
 at $2$ GeV). At $\beta = 5.8$ the lattice 
volume is $32^3 \times 64$ and the renormalized physical quark mass ranges 
from $15$ to $75$ MeV. The lattice spacings are determined using 
nucleon mass to pion mass ratio and
Sommer method. These determinations agree for the value of Sommer parameter 
$r_0= 0.44$ fm.    
The lattice spacings at
$\beta =5.6$ and $5.8$ are $0.069$ and $0.053$ fm respectively. 
The number of thermalized configurations ranges from $7000$ to $14000$ 
and the number of measured configurations ranges from $200$ to $500$.

The topological susceptibility $\chi=\int d^4x~C(r)$ 
with the TCDC, $C(r)=\langle q(x)q(0)\rangle, ~~ r=|x|$
where $q(x)$ is the topological charge density.
For $q(x)$, we use the lattice 
approximation developed for 
$SU(2)$ \cite{degrand},
modified for
$SU(3)$ \cite{hasenfratz1} and implemented in
the MILC code \cite{milc}. 
Unless otherwise stated we have used $3$ smearing steps\cite{hasenfratz2}  
in all our calculations.  

\section{Results}
\begin{figure}
\subfigure{
 \includegraphics[width=2.8in,clip]{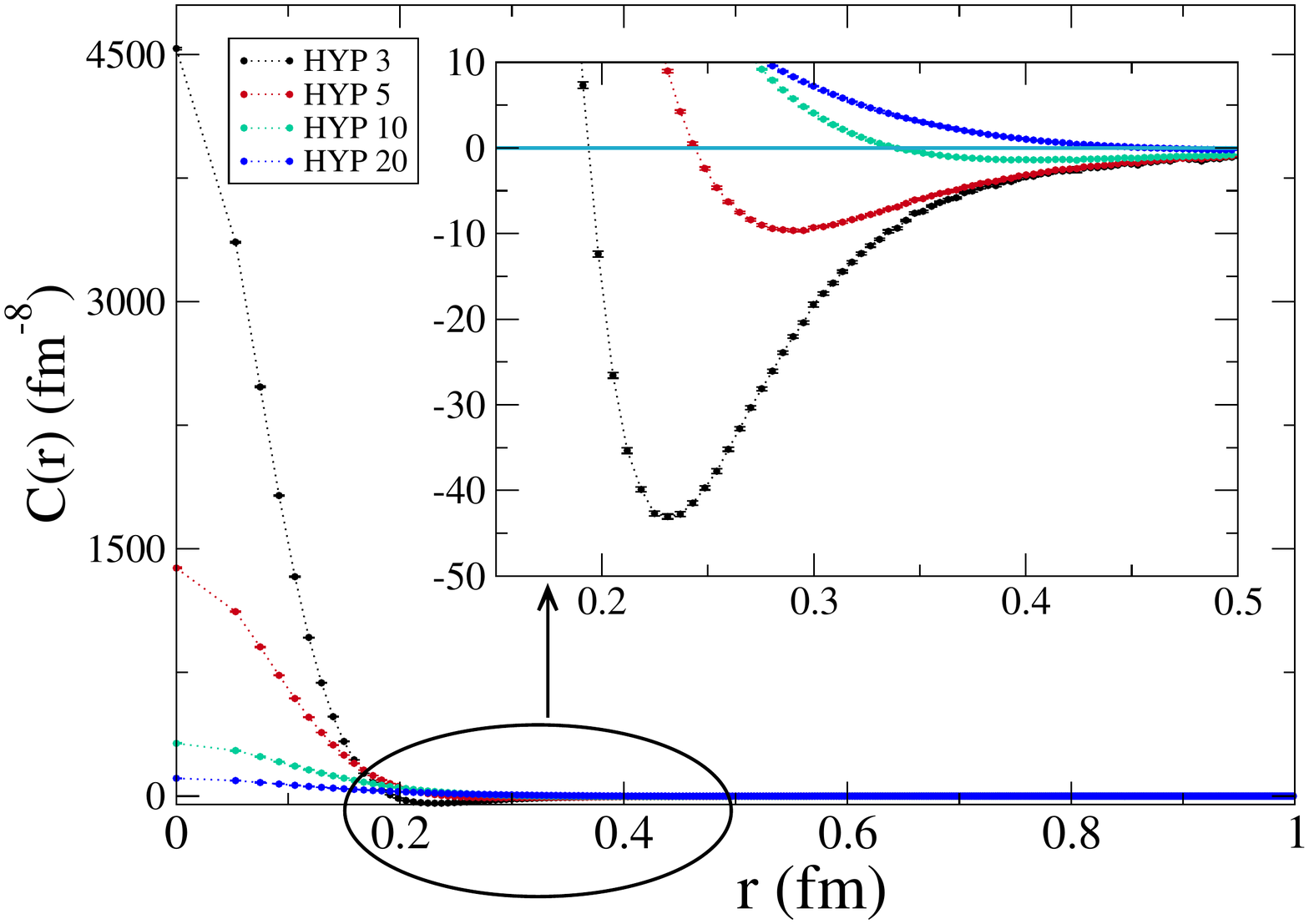}}
\subfigure{
 \includegraphics[width=2.8in,clip]{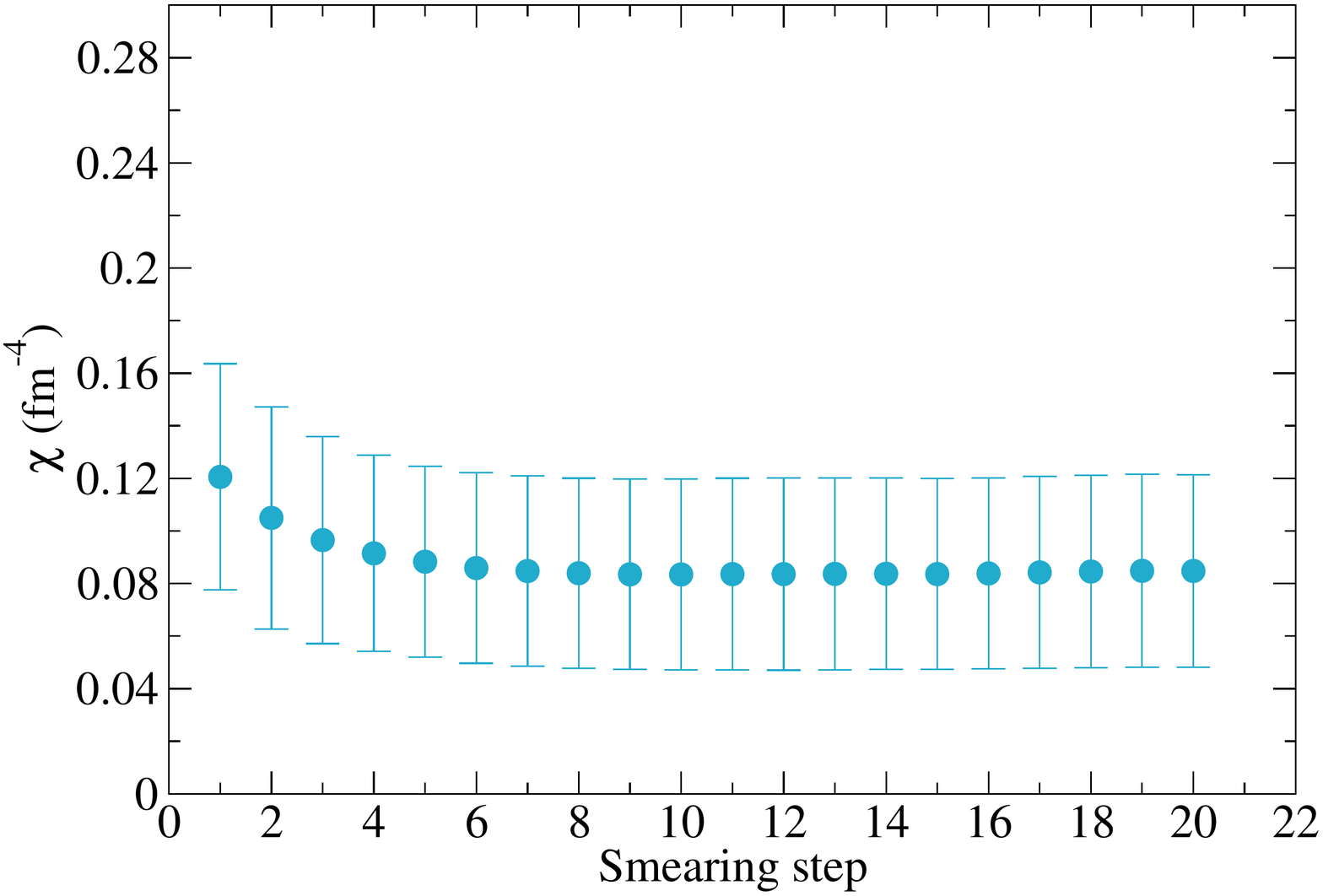}}
\vspace{-.5cm}
\caption{(left) Effect of smearing on $C(r)$ at $\beta=5.8$ and $\kappa=0.15462$ at lattice volume $32^3 \times 64$. 
(right) Effect of smearing on the $\chi$ at $\beta=5.8$ 
and $\kappa=0.15462$ at
lattice volume $32^3 \times 64$ (taken from Ref. \cite{ac2}).}
\label{fig9}
\end{figure}

In order to extract the topological charge density reliably on the lattice, 
using the algebraic definition, smearing of link field is essential. Smearing 
however smoothens out short distance singularities. 
Excessive smearing may in fact 
wipe out the fine details of the singularity structure. Both the positive and 
negative contributions to $\chi$ are affected in this manner. This is 
illustrated in Fig. \ref{fig9} (left)  where we show the effect of 
3, 5, 10 and 20 HYP smearing steps on $C(r)$ at $\beta=5.8$, $\kappa=0.15462$ and
lattice volume $32^3 \times 64$. However the 
susceptibility is remarkably stable
under smearing after three smearing steps as illustrated in
Fig. \ref{fig9} (right) (taken from Ref. \cite{ac2}).

\begin{figure}
 \subfigure{\includegraphics[width=2.8in,clip]
{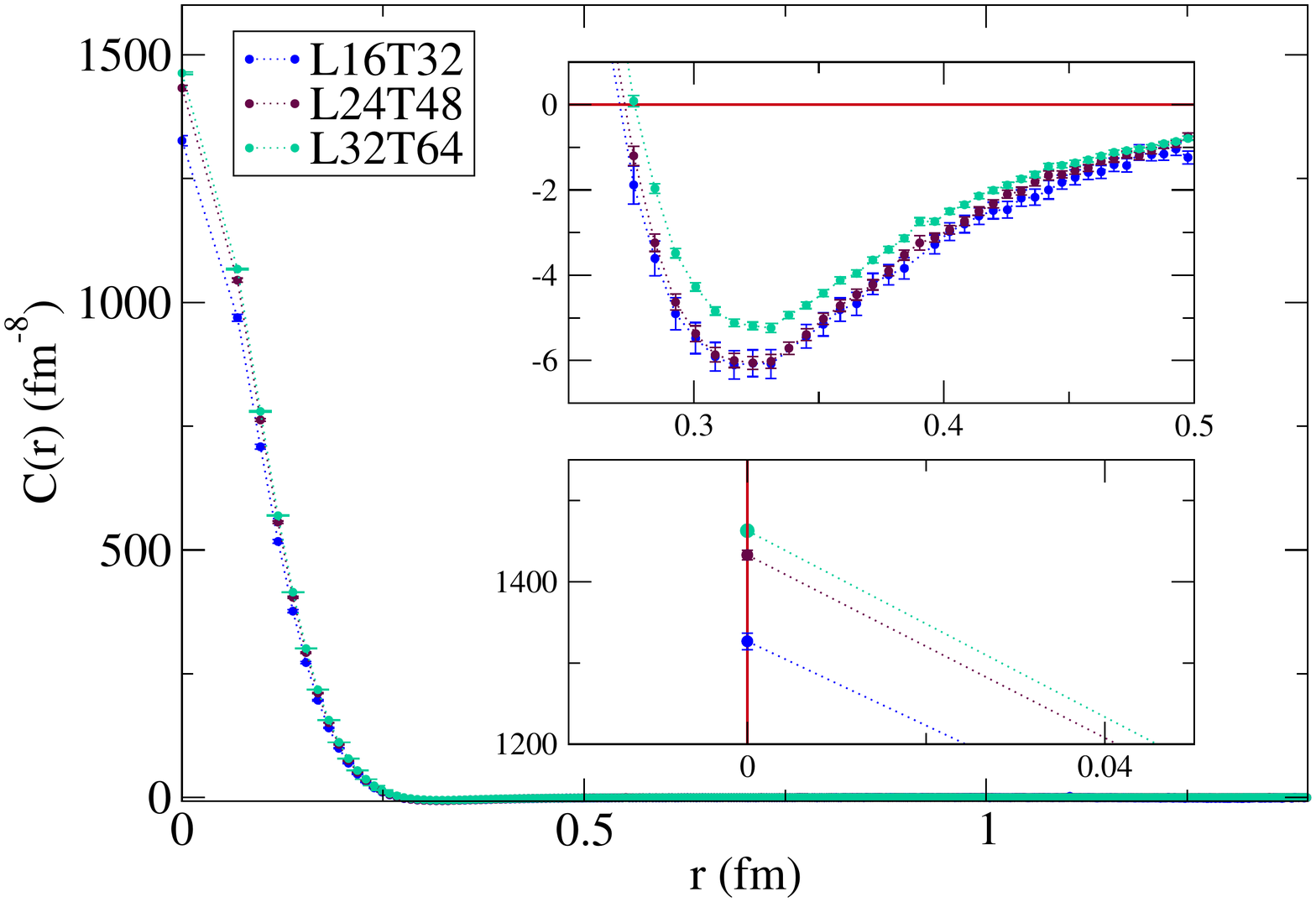}}
\subfigure{\includegraphics[width=2.8in,clip]
{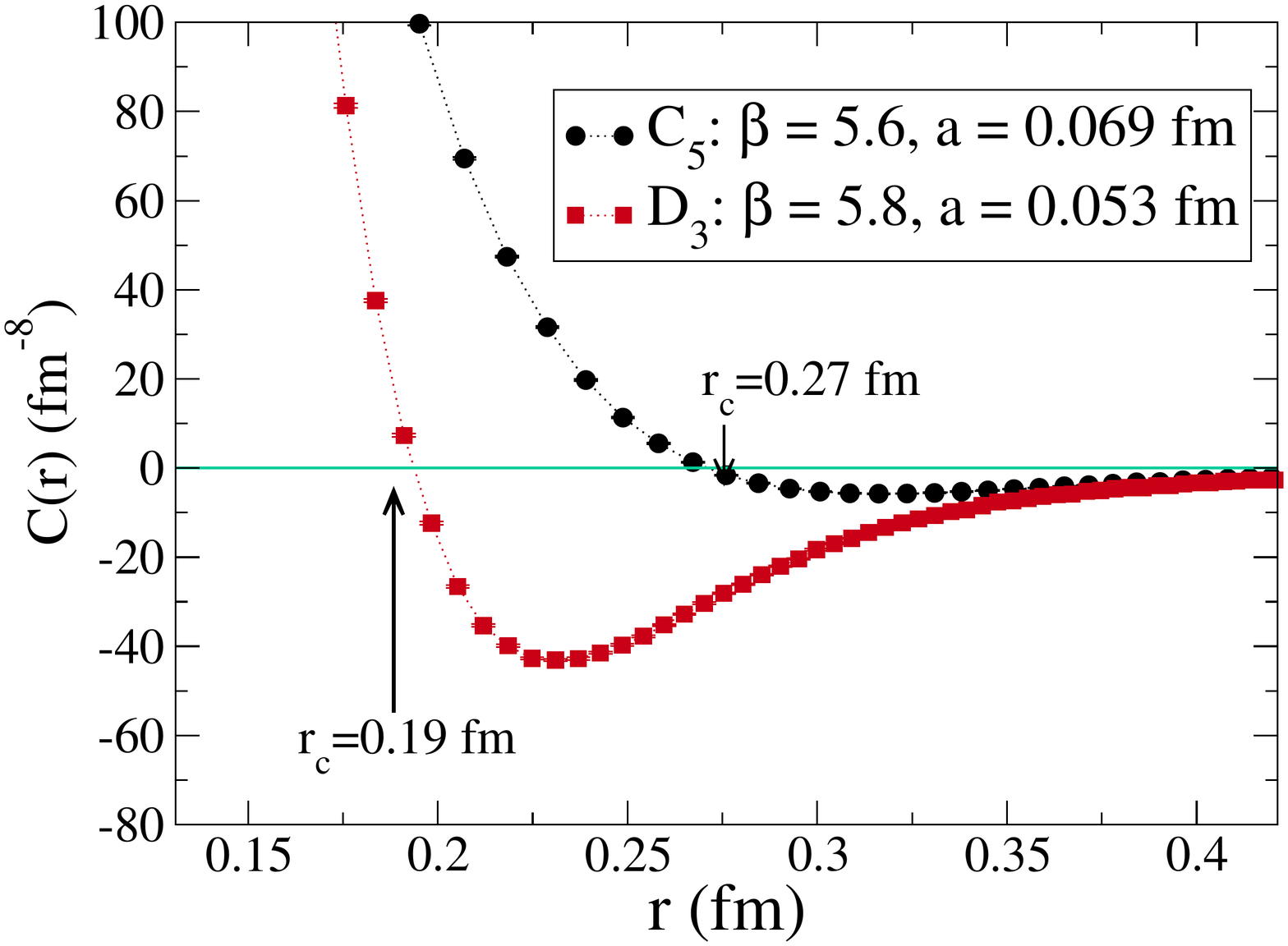}}
\vspace{-.5cm}
\caption{(left) Finite volume dependence of the $C(r)$  at $\beta=5.6$
and $\kappa=0.158$ at lattice volumes $16^3 \times 32$, $24^3 \times 48$ and
$32^3 \times 64$. (right)Comparison of the radius of the positive core of $C(r)$ at two 
different lattice spacings for comparable pion mass. 
Lattice volume is $32^3 \times 64$. }
\label{fig11}
\end{figure}
As was already stated, from theoretical considerations, we expect 
suppression of $\chi$ with decreasing volume at a fixed quark mass. 
In Fig. \ref{fig11} (left) we present the finite volume dependence of the $C(r)$  
at $\beta=5.6$ 
and $\kappa=0.158$ at lattice volumes $16^3 \times 32$, $24^3 \times 48$ and
$32^3 \times 64$. We find that as volume decreases, the magnitude of the 
contact term and radius of the positive core decrease and
the magnitude of the negative peak increases 
resulting in the 
suppression of topological susceptibility as volume decreases.



In Fig. \ref{fig11} (right), we compare the radius of the positive core of $C(r)$
at $\beta=5.6$ and $5.8$ for comparable
pion masses in physical units. The lattice volume is $32^3 \times 64$.
The figure clearly exhibits the shrinking of the 
radius of the positive core of $C(r)$ in physical units as one approaches 
the continuum.


\begin{figure}
\subfigure{
 \includegraphics[width=2.5in,clip]{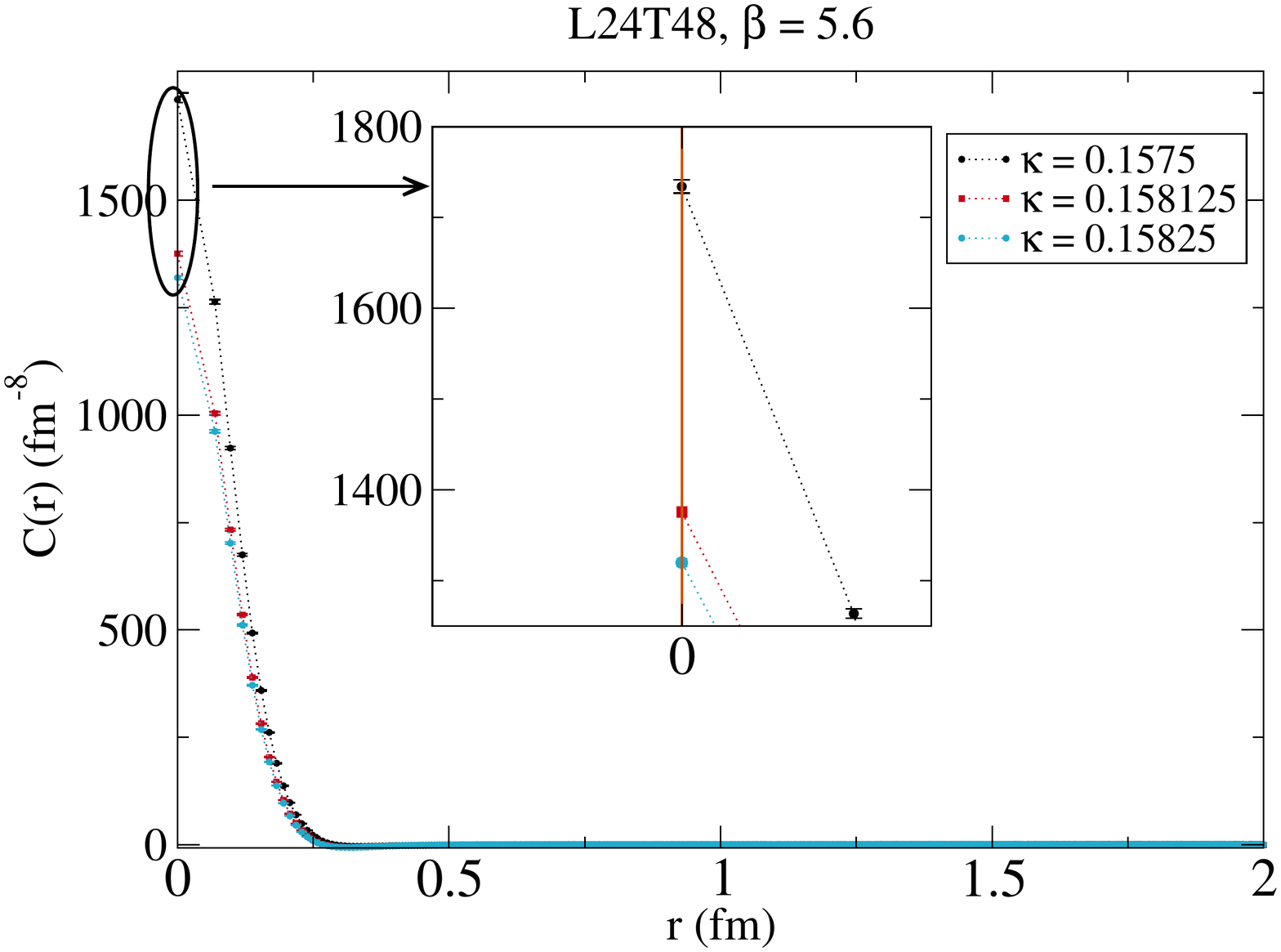}}
\subfigure{
 \includegraphics[width=2.5in,clip]{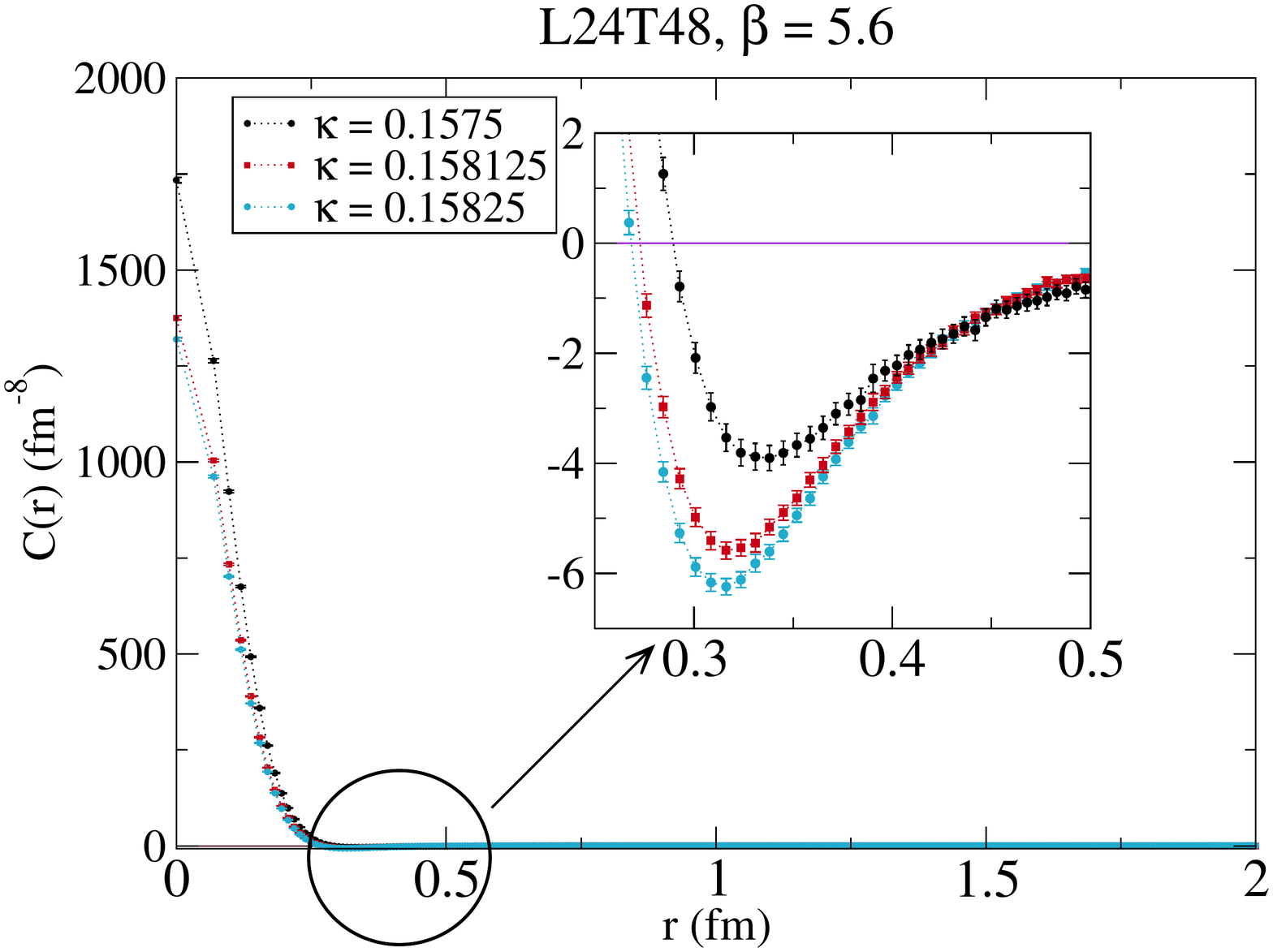}}
\vspace{-.5cm}
\caption{ The quark mass dependence of $C(r)$ 
with emphasis 
on the positive region (left) and with emphasis 
on the crossover from positive to the negative region of $C(r)$ 
and the negative peak  region (right)
at $\beta=5.6$ and lattice volume 
$24^3 \times 48$.}
\label{fig3}
\end{figure}

In order to understand the detailed mechanism behind the suppression 
of topological susceptibility with decreasing quark mass, we need to 
investigate the quark mass dependence of the various features of the $C(r)$.
In Fig. \ref{fig3} (left) we present the quark mass dependence of $C(r)$ 
with emphasis 
on the positive region at $\beta=5.6$ and lattice volume 
$24^3 \times 48$. The magnitude of the contact term $C(0)$ is seen to 
decrease with decreasing quark mass.   
In Fig. \ref{fig3} (right) we present the quark mass dependence of $C(r)$ 
with emphasis 
on the crossover from positive to the negative region of $C(r)$ and 
the negative peak  region at $\beta=5.6$ and lattice 
volume 
$24^3 \times 48$. The radius of the positive core and the 
magnitude of the negative peak of  $C(r)$ are seen to decrease and  
increase respectively with decreasing quark mass. The features presented 
in Figs. \ref{fig3} result 
in the suppression of the topological susceptibility with decreasing quark 
mass. MILC collaboration \cite{ab} has made a similar observation 
regarding the 
dependence of the negative peak on quark mass.  
\begin{figure}
\subfigure{
 \includegraphics[width=2.5in,clip]{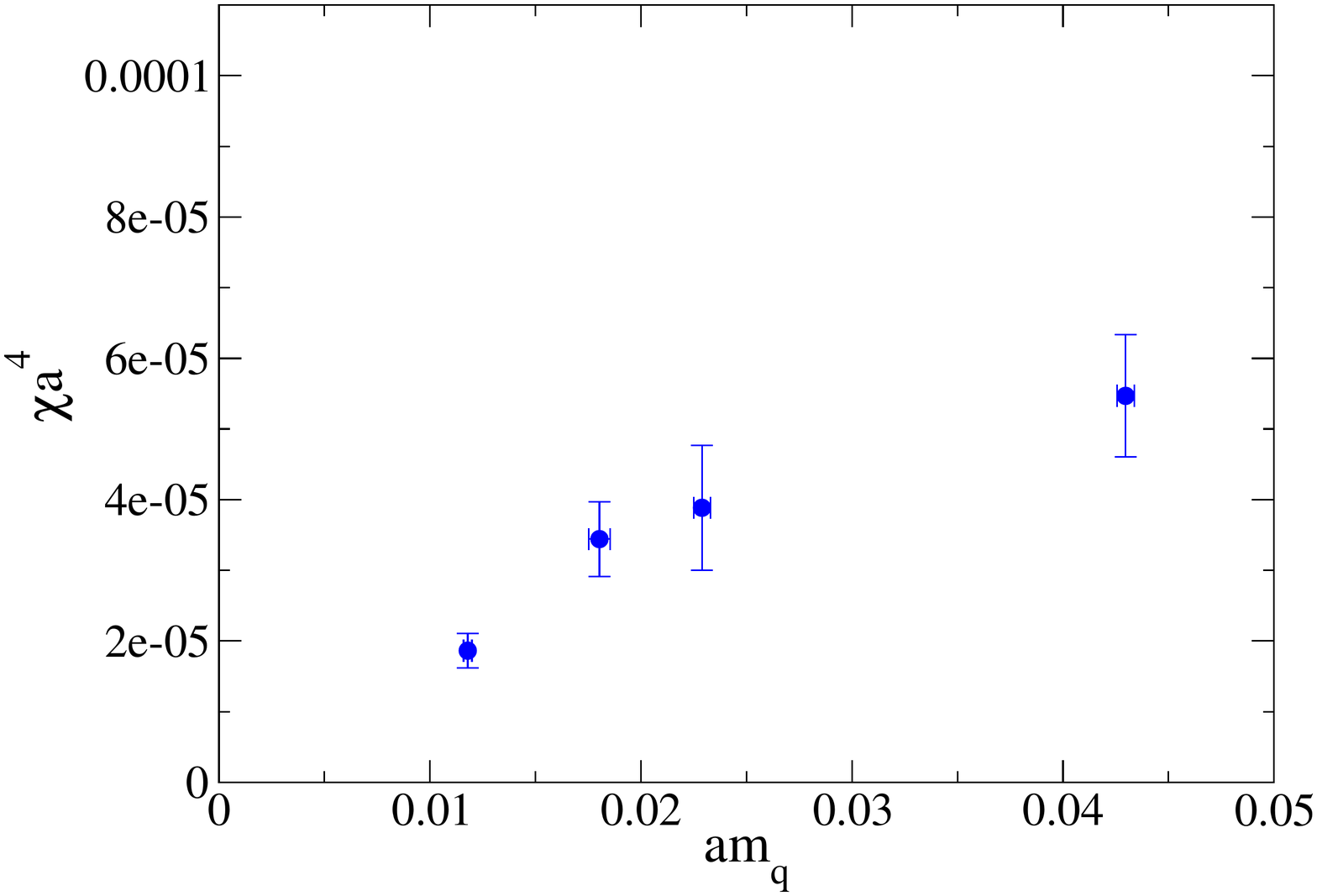}}
\subfigure{
 \includegraphics[width=2.5in,clip]{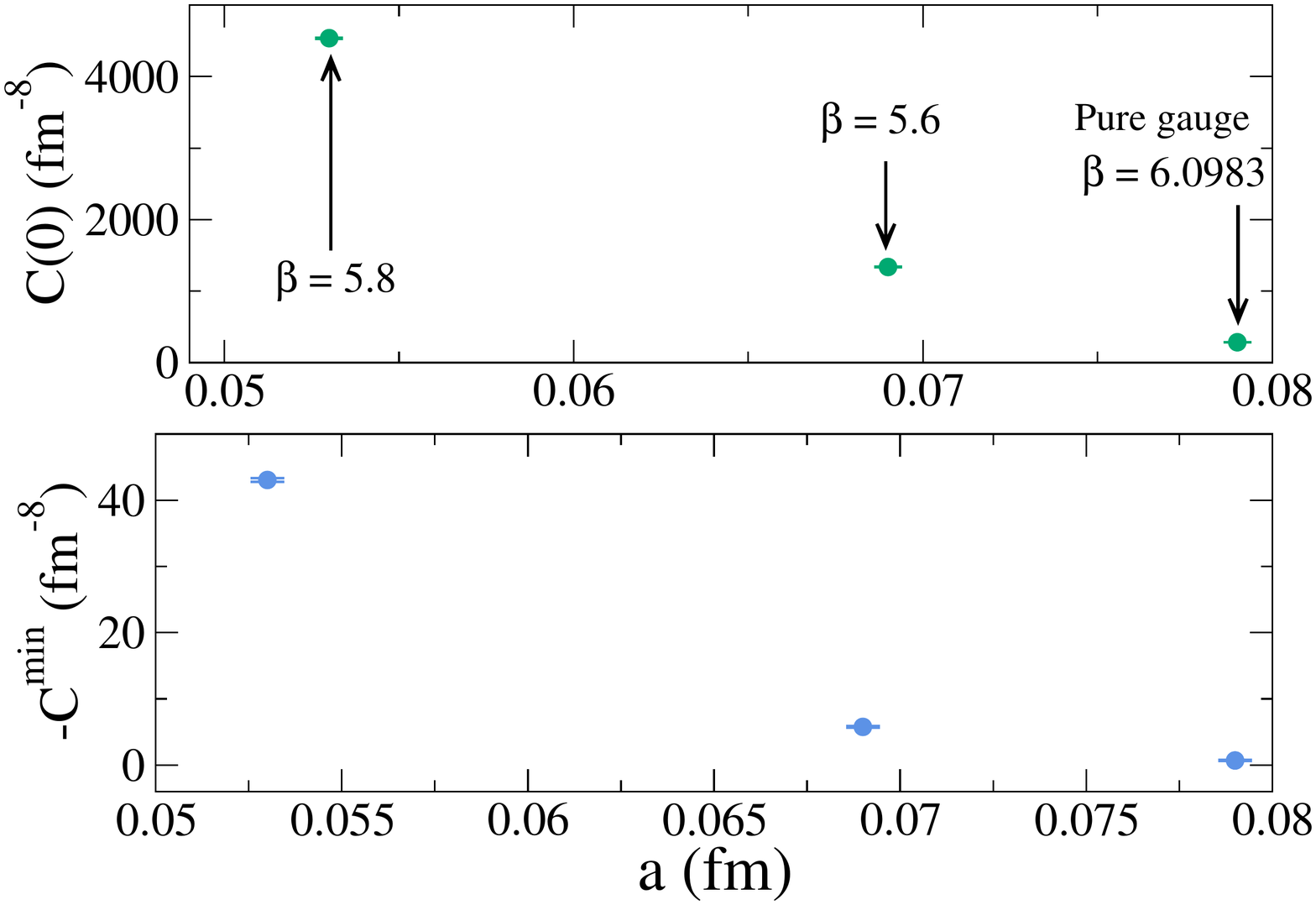}}
\vspace{-.5cm}
\caption{(left) Topological susceptibility at  
$\beta=5.6$ and lattice volume $24^3 \times 48$ as a function of the quark mass 
mass. (right) Lattice spacing dependence of the contact term 
and the negative peak of $C(r)$ at comparable pion mass for
$\beta=5.6$ and 5.8 and lattice volume $32^3 \times 64$. For comparison, 
the corresponding quantities for pure gauge lattice theory at $\beta=6.0983$ 
($a=0.078$fm) and lattice volume $24^3 \times 48 $ are also shown. }
\label{fig5}
\end{figure}
In Fig. \ref{fig5} (left) we present the corresponding topological susceptibilities 
($\beta=5.6$ and lattice volume $24^3 \times 48$) as a function of the quark 
mass which clearly shows the suppression as quark mass decreases.
This figure includes $\kappa=0.158$ in addition to the $\kappa$'s presented in Figs.
\ref{fig3}.

In Fig. \ref{fig5} (right) we show the lattice spacing dependence of the contact term 
and the negative peak of $C(r)$ at comparable pion mass in physical units for
$\beta=5.6$ and 5.8 and lattice volume $32^3 \times 64$. For comparison, 
the corresponding quantities for pure gauge lattice theory at $\beta=6.0983$ 
($a=0.078$fm) and lattice volume $24^3 \times 48 $ are also shown. Both the  
contact 
term 
and the negative peak of $C(r)$ increase with decreasing lattice spacing,
in accordance  with the expectation from the continuum theory.   

\begin{figure}
\subfigure{
 \includegraphics[width=2.5in,clip]{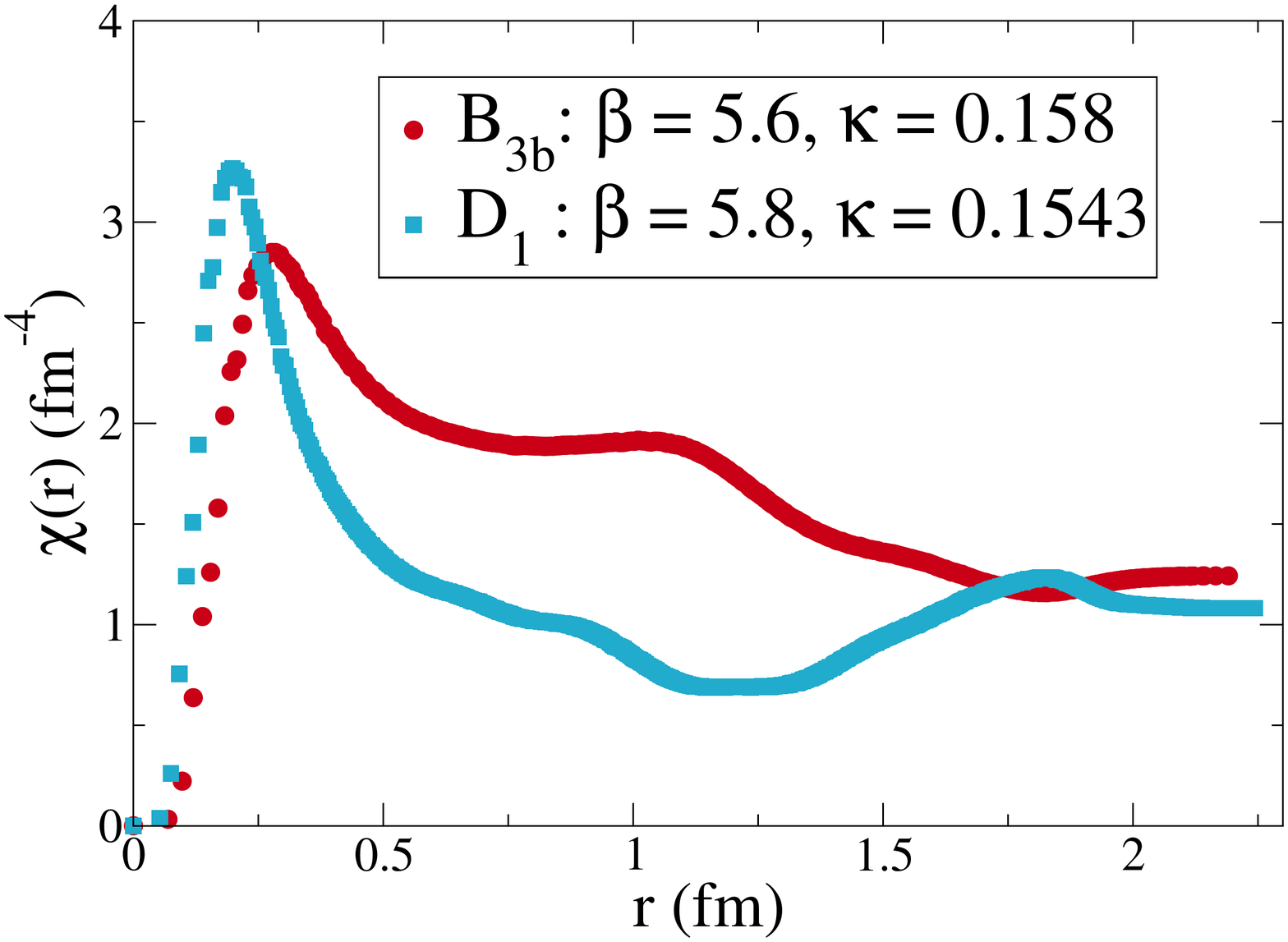}}
\subfigure{
 \includegraphics[width=2.5in,clip]{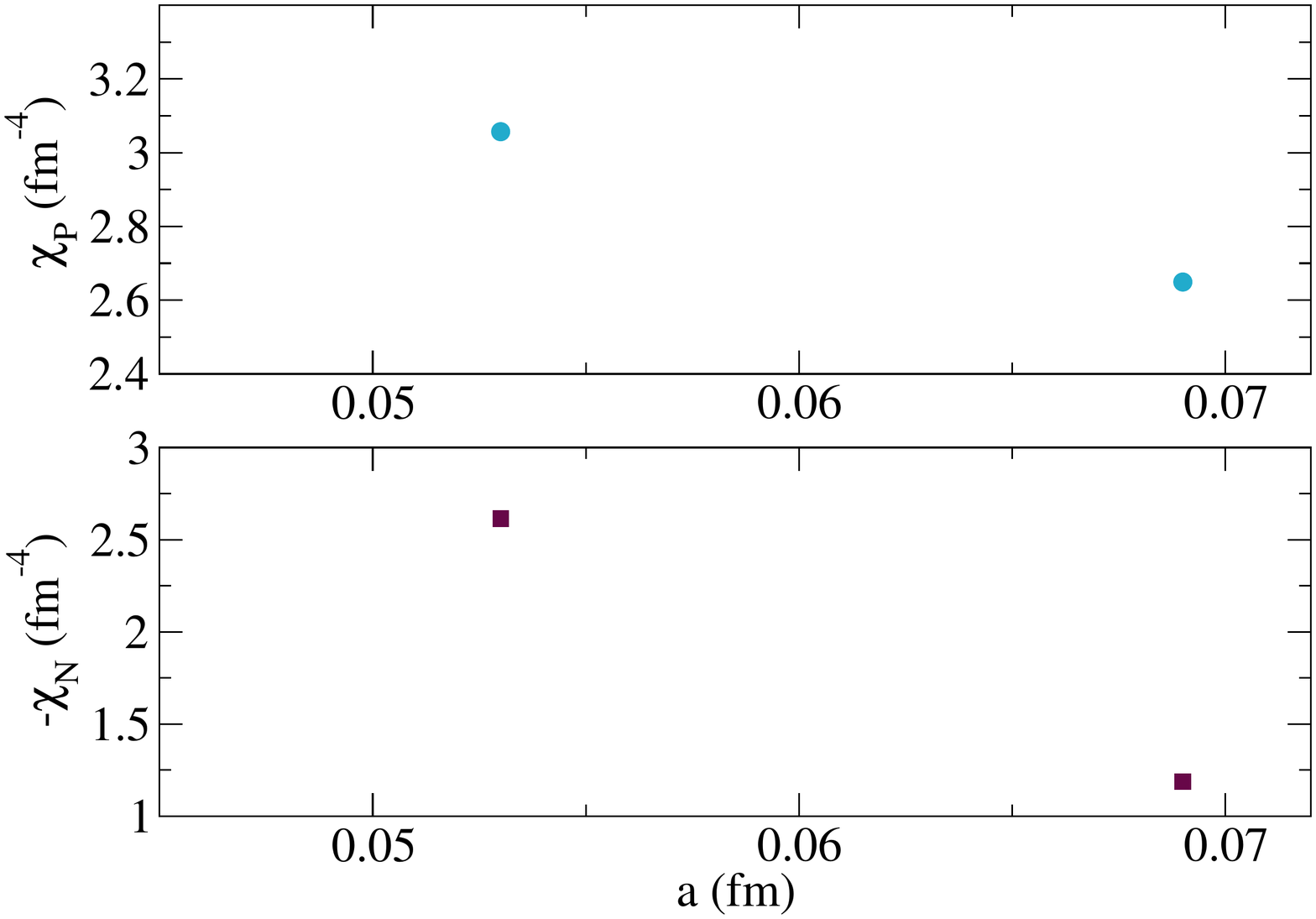}}
\vspace{-.5cm}
\caption{The function $\chi(r)$, defined in Eq. (3.1) as a function of 
$r$ (left) and contributions to the susceptibility from positive and negative 
parts of $C(r)$ (right)
at $\beta=5.6$ and $5.8$ at comparable pion masses.}
\label{fig7}
\end{figure}

From the definition of the topological susceptibility 
$\chi=\int d^4x ~C(r)=\int 2\pi^2(r^3) dr~C(r)$
it is useful to define \cite{ab} a {\em local susceptibility}
\be
\chi(r)=\int^r_0 2\pi^2(r'^3)dr'~ C(r')
\ee 
in order to exhibit the lattice spacing dependence more clearly. In 
Fig. \ref{fig7} (left) we plot $\chi(r)$ versus $r$ at two lattice spacings at 
comparable quark masses. The contribution from the positive part of $C(r)$ results in a 
peak at short distance. This is followed by a decrease due to the negative 
part of $C(r)$. As lattice spacing decreases, the contribution from the positive part
increases resulting in the increase of the peak of $\chi(r)$.     

Define the contributions to the susceptibility from the positive and negative 
parts 
of 
$C(r)$ as
$
\chi_P =\int^{r_c}_02\pi^2(r'^3) dr'~C(r')~~ {\rm and}~~
\chi_N =\int_{r_c}^{\infty}2\pi^2(r'^3)dr'~ C(r') \nonumber
$
respectively.
According to the expectations from continuum theory, the negative singularity 
close to the origin and the positive singularity at the origin are both 
nonintegrable. Thus the contributions to $\chi$ from positive and negative
parts of $C(r)$ are expected to diverge, nevertheless resulting in a finite 
$\chi$ due to cancellation. In Fig. \ref{fig7} (right), we plot the contributions to 
the susceptibility from 
positive and negative parts of $C(r)$ at $\beta=5.6$ and $5.8$ at comparable 
pion masses. The data exhibited in Fig. \ref{fig7} (right) are in accordance with these
expectations.


It is known that 
the topological susceptibility decreases with decreasing quark mass
and decreasing volume.
To understand the mechanisms leading to these suppressions,
in this work, we carry 
out a detailed study of the  two-point TCDC.
We have shown that, 
with naive Wilson fermion and gauge action, 
(1) the two-point TCDC is negative beyond a positive 
core and radius of the 
core shrinks 
as lattice spacing decreases, 
(2) as volume decreases, the magnitude of the 
contact term and the radius of the positive core decrease
and the magnitude of the negative peak increases 
resulting in the 
suppression of topological susceptibility as volume decreases, 
(3) the contact term and radius of the positive core decrease with
decreasing quark mass at a given lattice spacing and the
negative peak increases with decreasing quark mass resulting
in the suppression of the topological susceptibility with decreasing quark 
mass, 
(4)  increasing levels of smearing suppresses the contact term and
the negative 
peak keeping the susceptibility intact and 
(5) both the contact term and the negative peak diverge in nonintegrable 
fashion as lattice spacing decreases.  


\vskip .05in
{\bf Acknowledgements}
\vskip .05in
  Numerical calculations are carried out on Cray XD1 and Cray XT5 systems 
supported 
by the 10th and 11th Five Year Plan Projects of the Theory Division, SINP under
the DAE, Govt. of India. We thank Richard Chang for the prompt maintainance of 
the systems and the help in data management. This work was in part based on 
the public lattice gauge theory codes of the 
MILC collaboration \cite{milc} and  Martin L\"{u}scher \cite{ddhmc}.



\end{document}